\documentclass[12pt]{article}

\usepackage{graphicx}
\begin{document}

\begin{center}
{\bf Corrections to Reissner-Nordstr\"om black hole solution due to exponential nonlinear electrodynamics } \\
\vspace{5mm} S. I. Kruglov
\footnote{E-mail: serguei.krouglov@utoronto.ca}
\underline{}
\vspace{3mm}

\textit{Department of Chemical and Physical Sciences, University of Toronto,\\
3359 Mississauga Road North, Mississauga, Ontario L5L 1C6, Canada} \\
\vspace{5mm}
\end{center}

\begin{abstract}
An exponential model of nonlinear electrodynamics coupled with the gravitational field is analyzed. It was shown that weak, dominant,
and strong energy conditions are satisfied. We obtain corrections to Coulomb's law at
$r\rightarrow\infty$. It was demonstrated that electric-magnetic duality between ${\cal F}$ and ${\cal P}$  frames is violated.
The asymptotic black hole solution at $r\rightarrow \infty$ is found that contains corrections to the Reissner-Nordstr\"om solution.
The electromagnetic mass of the black hole is calculated, expressed through the parameter $\beta$ of the model.
\end{abstract}

\section{Introduction}

The problem of singularity of an electric field in the center of charged point-like particles and the problem of infinite electromagnetic energy can be solved within nonlinear electrodynamics (NLED). Maxwell's electrodynamics follows from NLED for weak fields, and therefore, it can be considered as an approximation. It was mentioned in \cite{Jackson} that for strong electromagnetic fields classical
electrodynamics should be modified because of the self-interaction of photons. One example of NLED is
Born-Infeld electrodynamics (BIE) \cite{Born} where an upper limit on the electric field exists at the center of charged particles. The total electric energy of point-like charges is also finite in the framework of BIE. Different models of NLED  were investigated in
\cite{Shabad} - \cite{Kruglov7} where problems of singularities were studied.

At the same time, NLED can be a source of the universe acceleration \cite{Garcia}-\cite{Kruglov4}. We note that electromagnetic fields in BIE do not drive the acceleration of the universe \cite{Novello1} and BIE has the problem of causality \cite{Quiros}.
The black hole solutions in General Relativity (GR), where NLED is the matter field, were investigated in \cite{Breton} - \cite{Kruglov8}.
Black hole solutions in these models approach to the Reissner–Nordstr\"{o}m (RN) solution at $r\rightarrow \infty$.

In this letter we obtain  the solution for charged black holes within NLED proposed in \cite{Kruglov0} and find corrections to the Coulomb law and to the RN black hole solution.

The paper is organized as follows. In section II we show that weak, dominant, and strong energy conditions are satisfied in our model.
NLED coupled with GR is considered in section III. Corrections to Coulomb's law are obtained.
It was demonstrated that electric-magnetic duality between ${\cal F}$ and ${\cal P}$  frames is broken. In section IV we find Black hole solution which
asymptotically approaches to RN solution. Electromagnetic mass of a black hole is calculated and metric function is obtained.
We find corrections to RN solution at $r\rightarrow\infty$. Section V is devoted to the conclusion.

The units $c=\hbar=1$ and the metric $\eta=\mbox{diag}(-1,1,1,1)$ are used.

\section{The model of nonlinear electrodynamics}

Here we consider exponential nonlinear electrodynamics proposed in \cite{Kruglov0}.
This type of NLED coupled with GR is of interest because it leads to accelerated expansion of universe driven by electromagnetic
fields. The Lagrangian density of exponential nonlinear electrodynamics is given by
\begin{equation}
{\cal L} = -{\cal F}\exp(-\beta{\cal F}),
 \label{1}
\end{equation}
where the parameter $\beta$ has the dimension of the (length)$^4$, and  ${\cal F}=(1/4)F_{\mu\nu}F^{\mu\nu}=(\textbf{B}^2-\textbf{E}^2)/2$
($F_{\mu\nu}=\partial_\mu A_\nu-\partial_\nu A_\mu$ is the electromagnetic field strength tensor).
The upper bound on the parameter $\beta$ ($\beta \leq 1\times 10^{-23}$ T$^{-2}$) was obtained in \cite{Kruglov0} from PVLAS experiment.
The symmetric energy-momentum tensor is written as \cite{Kruglov0}:
\begin{equation}
T^{\mu\nu}=\exp(-\beta{\cal F})\left[\left(\beta {\cal F}-1\right)F^{\mu\lambda}F^\nu_{~\lambda}+g^{\mu\nu}{\cal F}\right]
\label{2}
\end{equation}
with the trace
\begin{equation}
{\cal T}\equiv T_{\mu}^{~\mu}=4\beta {\cal F}^2\exp(-\beta{\cal F}).
\label{3}
\end{equation}
Maxwell's electrodynamics is recovered at $\beta\rightarrow 0$ and the trace becomes zero.
The scale invariance and dilatation invariance are violated because of the dimensional parameter $\beta$.
Equations of motion (field equations) which follow from the Lagrangian density (1) are nonlinear gauge-invariant equations with higher derivatives. Therefore, photons in our model are massless unless the gauge ($U(1)$ group) symmetry holds. Due to nonlinear interaction of electromagnetic
fields, the gauge symmetry possibly could be spontaneously broken and, for example, condensate $\langle A_\mu A^\mu\rangle$ will become not-zero,
$\langle A_\mu A^\mu\rangle\neq 0$. In this case photons can acquire a mass. Thus, vector $Z$ and $W^\pm$ bosons become massive after
the interaction with the scalar Higgs boson that
 is described by nonlinear scalar fields.
Fermion fields also can acquire masses because of self-interaction (see, for example, \cite{Kruglov9}). The investigation of possibility of
spontaneous violation of gage symmetry in our model, described by Eq. (1), is beyond of the scope of this paper.

Now we investigate the energy conditions which are important for viability of models.
Consider the weak energy condition (WEC) \cite{Hawking} that guarantees that the energy density is positive for any local observer,
\begin{equation}
\rho\geq 0,~~~\rho+p^m\geq 0 ~~~~ (m=1,~2,~3),
\label{4}
\end{equation}
$\rho$ being the energy density and $p^m$ are principal pressures. We obtain from Eq. (2)
\begin{equation}
\rho=T_0^{~0}=\left(\frac{B^2+E^2}{2}-\beta E^2{\cal F}\right)\exp(-\beta {\cal F}),
\label{5}
\end{equation}
\begin{equation}
p^m=-T_m^{~m}=\left[(1-\beta{\cal F})(B^2-E^mE_m-B^mB_m)-{\cal F}\right]\exp(-\beta {\cal F}),
\label{6}
\end{equation}
$m=1,2,3$ and there is not the summation in the index $m$ in Eq. (6).
If $\textbf{B}=0$, we obtain from Eqs. (5) and (6) the energy density and principal pressures for the presence of pure electrical field
\begin{equation}
\rho_E=\frac{E^2}{2}(1+\beta E^2)\exp(\beta E^2/2),                                                                                                                                                                                                       \label{5}
\label{7}
\end{equation}
\begin{equation}
p^m_E=\left[\frac{E^2}{2}-\left(1+\frac{\beta E^2}{2}\right) E^mE_m\right]\exp(\beta E^2/2)~~~~(m=1,2,3),
\label{8}
\end{equation}
It follows from Eq. (7) that $\rho_E\geq 0$. From Eqs. (7) and (8), we find
\begin{equation}
\rho_E+p^m_E=\left(1+\frac{\beta E^2}{2}\right)\left(E^2-E^m E_m\right)\exp(\beta E^2/2)\geq 0~~~~(m=1,2,3).
\label{9}
\end{equation}
As a result, WEC holds for any electric field values.
For the case $\textbf{E}=0$, $\textbf{B}\neq 0$, one obtains from Eqs. (5) and (6)
\begin{equation}
\rho_M=\frac{B^2}{2}\exp(-\beta B^2/2),
\label{10}
\end{equation}
\begin{equation}
p^m_M=\left[\frac{B^2}{2}-B_mB^m-\frac{\beta B^2}{2}(B^2- B^mB_m)\right]\exp(-\beta B^2/2).
\label{11}
\end{equation}
We see from Eq. (10) that $\rho_B\geq 0$. By virtue of Eqs. (10) and (11) one finds
\begin{equation}
\rho_M+p^m_M=\left(1-\frac{\beta B^2}{2}\right)\left(B^2-B^m B_m\right)\exp(-\beta B^2/2).
\label{12}
\end{equation}
Thus, WEC is satisfied for the case $\textbf{E}=0$, $\textbf{B}\neq 0$ if $\beta B^2\leq 2$.

The dominant energy condition (DEC) \cite{Hawking} is given by the relations
\begin{equation}
\rho\geq 0,~~~\rho+p^m\geq 0,~~~\rho-p^m\geq 0~~(m=1,~2,~3).
\label{13}
\end{equation}
From Eqs. (7) and (8) at $\textbf{B}=0$, $\textbf{E}\neq 0$ we obtain
\begin{equation}
\rho_E-p^m_E=\left[\frac{\beta E^2}{2}+\left(1+\frac{\beta E^2}{2}\right) E^mE_m\right]\exp(\beta E^2/2)\geq 0.
\label{14}
\end{equation}
As a result, DEC is satisfied for this case and shows that the speed of sound is less than the speed of light.
If $\textbf{E}=0$, $\textbf{B}\neq 0$, one finds from Eqs. (10) and (11)
\begin{equation}
\rho_M-p^m_M=\left[\frac{\beta B^2}{2}(B^2-B_mB^m)+ B^mB_m\right]\exp(-\beta B^2/2)\geq 0,
\label{15}
\end{equation}
and DEC is satisfied.

The strong energy condition (SEC) \cite{Hawking} is defined as
\begin{equation}
\rho+\sum_{m=1}^3p^m\geq 0.
\label{16}
\end{equation}
By virtue of Eqs. (7) and (8), one obtains
\begin{equation}
\rho_E+\sum_{m=1}^3p^m_E=E^2\exp(\beta E^2/2)\geq 0,
\label{17}
\end{equation}
and, as a result, SEC holds.

For the case $\textbf{E}=0$, $\textbf{B}\neq 0$, we find from Eqs. (10) and (11)
\begin{equation}
\rho_M+\sum_{m=1}^3p^m_M=B^2(1-\beta B^2)\exp(-\beta B^2/2),
\label{18}
\end{equation}
and, therefore, SEC occurs at $B\leq 1/\sqrt{\beta}$.

The pressure is given by
\begin{equation}
p={\cal L}+\frac{E^2-2B^2}{3}{\cal L}_{\cal F}=\left[\frac{2B^2-E^2}{3}(1-\beta {\cal F})-{\cal F}\right]\exp(-\beta {\cal F})=\frac{1}{3}\sum_{m=1}^3p^m,
\label{19}
\end{equation}
where $ {\cal L}_{\cal F}=\partial {\cal L}/\partial {\cal F}$.
Eq. (16) may be written in the form $\rho +3p\geq 0$ which due to Friedmann's equation shows that electrically charged universe decelerates. But
magnetically charged universe accelerates in our model at $B> 1/\sqrt{\beta}$ \cite{Kruglov0} that follows from Eq. (18).

\section{NLED coupled with GR and black holes}

Let us consider the action for the case of NLED coupled with GR
\begin{equation}
S=\int d^4x\sqrt{-g}\left[\frac{1}{2\kappa^2}R+ {\cal L}\right],
\label{20}
\end{equation}
where $\kappa^2=8\pi G\equiv M_{Pl}^{-2}$, $M_{Pl}$ is the reduced Planck mass, $G$ is the Newton constant, and $R$ is the Ricci scalar. Varying action (20), we obtain the Einstein and NLED equations
\begin{equation}
R_{\mu\nu}-\frac{1}{2}g_{\mu\nu}R=-\kappa^2T_{\mu\nu},
\label{21}
\end{equation}
\begin{equation}
\partial_\mu\left[\sqrt{-g}F^{\mu\nu}\left(1-\beta{\cal F}\right)\exp(-\beta{\cal F})\right]=0.
\label{22}
\end{equation}
To obtain the static charged black hole solutions to Eqs. (21) and (22) we consider the spherically symmetric line element in
$(3+1)$-dimensional spacetime
\begin{equation}
ds^2=-f(r)dt^2+\frac{1}{f(r)}dr^2+r^2(d\vartheta^2+\sin^2\vartheta d\phi^2).
\label{23}
\end{equation}
At $\textbf{B}=0$ for the metric (23) and with the help of the relation ${\cal F}=-[E(r)]^2/2$ (the vector-potential has non-zero component $A_0(r)$), from Eq. (22) one obtains
\begin{equation}
\partial_r\left[r^2 E(r)\left(1+\frac{\beta [E(r)]^2}{2}\right)\exp(\beta [E(r)]^2/2)\right]=0.
\label{24}
\end{equation}
After integration of Eq. (24), we find
\begin{equation}
E(r)\left(1+\frac{\beta [E(r)]^2}{2}\right)\exp(\beta [E(r)]^2/2)=\frac{Q}{r^2},
\label{25}
\end{equation}
where $Q$ is the constant of integration. Let us introduce the dimensionless variables
\begin{equation}
x=\frac{r}{\sqrt{Q}}\left(\frac{2}{\beta}\right)^{1/4},~~~~y=E(r)\sqrt{\frac{\beta}{2}}.
\label{26}
\end{equation}
By virtue of variables (26), Eq. (25) is written as
\begin{equation}
y\left(1+y^2\right)\exp(y^2)=\frac{1}{x^2}.
\label{27}
\end{equation}
The function $y(x)$ is represented in Fig. 1.
\begin{figure}[h]
\includegraphics[height=3.0in,width=3.0in]{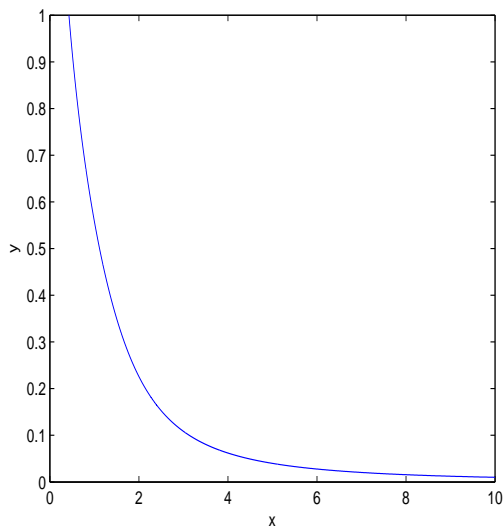}
\caption{\label{fig.1}The function $y(x)$.}
\end{figure}
At $x\rightarrow 0$ one has $y\rightarrow \infty$.
We obtain the asymptotic behavior of the function $y(x)$ (27) at $x\rightarrow\infty$
\begin{equation}
y=\frac{1}{x^2}-\frac{2}{x^6}+\frac{12}{x^{10}}+{\cal O}(x^{-13}).
\label{28}
\end{equation}
At $r\rightarrow \infty$ one finds the asymptotic value of the electric field from Eqs. (26) and (28)
\begin{equation}
E(r)=\frac{Q}{r^2}-\beta\left(\frac{ Q^{3}}{r^6}-\frac{3\beta Q^5}{r^{10}}+{\cal O}(r^{-13})\right).
\label{29}
\end{equation}
Thus, the $Q$ is the charge and Eq. (29) gives the corrections to Coulomb's law at $r\rightarrow\infty$. Corrections are in the order of $r^{-6}$, and at $\beta=0$ we come to the Coulomb law $E=Q/r^2$. It should be mentioned that corrections to Coulomb's law in the order of $r^{-6}$ were obtained in other models of NLED \cite{Hendi} (see also \cite{Kruglov6}, \cite{Kruglov8}).
After integrating function (29), one obtains, using $A_0(r)=-\int E(r)dr$, the asymptotic value of the electric potential at $r\rightarrow\infty$
\begin{equation}
A_0(r)=\frac{Q}{r}-\beta\left(\frac{ Q^{3}}{5r^5}-\frac{3\beta Q^5}{9r^9}+{\cal O}(r^{-12})\right).
\label{30}
\end{equation}
It has to be mentioned that massive photons are described by the Proca equation and a static electric potential is given by the Yukawa (or Debye) potential (see, for example \cite{Tu})
\begin{equation}
\phi(r)\equiv A_0=\frac{Q}{r}\exp(-m_\gamma r),
\label{31}
\end{equation}
where $m_\gamma$ is a photon mass. The electric field corresponding to the potential (31) is
\begin{equation}
E(r)=\left(\frac{Q}{r^2}+\frac{Qm_\gamma}{r}\right)\exp(-m_\gamma r).
\label{32}
\end{equation}
The upper limit on the rest mass of a photon is $m_\gamma\leq 1\times 10^{-49}$ g $\equiv 6\times 10^{-17}$ eV \cite{Eidelman}.
The behaviour of the electric field (32) at $r\rightarrow \infty$ is different from (29) due to damping by the exponential factor.

\subsection{Electric-magnetic duality}

With the help of a Legendre transformation \cite{Garcia1} one comes to another form of NLED within P-framework. Consider the tensor
\begin{equation}
P_{\mu\nu}=\frac{1}{2}F_{\mu\nu}{\cal L}_{\cal F} =-\frac{1}{2}F_{\mu\nu}\left(1-\beta {\cal F} \right)\exp(-\beta {\cal F}).
\label{33}
\end{equation}
Then by virtue of Eq. (33) we obtain the invariant
\begin{equation}
P=P_{\mu\nu}P^{\mu\nu}={\cal F}\left(1-\beta {\cal F} \right)^2 \exp(-2\beta {\cal F}).
\label{34}
\end{equation}
We introduce the Hamilton-like variable
 \begin{equation}
{\cal H}=2{\cal F}{\cal L}_{\cal F}-{\cal L}=-{\cal F}\left(1-2\beta {\cal F} \right) \exp(-\beta {\cal F}).
\label{35}
\end{equation}
It follows from Eqs. (35) and (7) that ${\cal H}=\rho_E$ (at \textbf{B}=0) i.e. it is the electric energy density. The relations
\begin{equation}
{\cal L}_{\cal F}{\cal H}_P=1,~~~~P{\cal H}_P^2={\cal F},~~~~{\cal L}=2P{\cal H}_P-{\cal H}
\label{36}
\end{equation}
hold, where
\begin{equation}
{\cal H}_P=\frac{\partial {\cal H}}{\partial P}=-\frac{[1-5\beta {\cal F}+2(\beta {\cal F})^2]\exp(\beta {\cal F})}{1-6\beta {\cal F}+7(\beta {\cal F})^2-2(\beta {\cal F})^3}.
\label{37}
\end{equation}
The plot of the function $P({\cal F})$ is represented in Fig. 2. which shows that the function ${\cal F}(P)$ is not a monotonic function.
\begin{figure}[h]
\includegraphics[height=3.0in,width=3.0in]{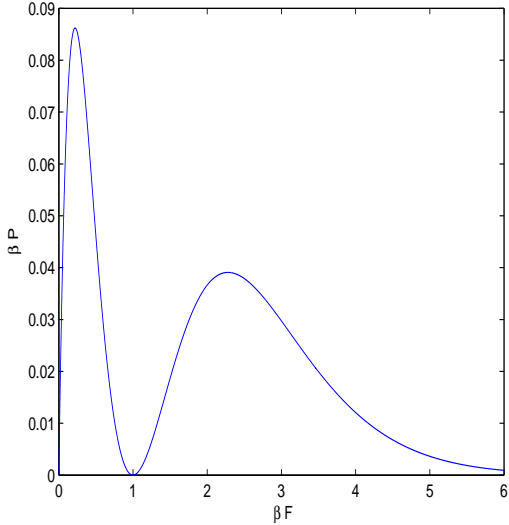}
\caption{\label{fig.2}The function $\beta P$ vs $\beta {\cal F}$.}
\end{figure}

Therefore, there is not a one to one correspondence between ${\cal F}$ and $P$ frames \cite{Bronnikov},
and an exact electric-magnetic duality between these frames does not hold.
At $\beta{\cal F}\ll 1$, for weak fields, both models (1) and (35) are converted to the Maxwell theory, ${\cal L}={\cal H}=-{\cal F}$.

\section{Asymptotic Reissner-Nordstr\"{o}m black hole solution}

 From Eq. (21) one obtains the Ricci curvature
\begin{equation}
R=\kappa^2{\cal T}.
\label{38}
\end{equation}
By virtue of Eqs. (3) and (38) at $\textbf{B}=0$, we find the Ricci scalar
\begin{equation}
R=\kappa^2\beta[E(r)]^4\exp(\beta[E(r)]^2/2).
\label{39}
\end{equation}
The electric field at $r\rightarrow \infty$ in accordance with Eq. (29) approaches to zero, $E(r)\rightarrow 0$. As a result, the Ricci scalar
at $r\rightarrow \infty$ goes to zero, $R \rightarrow 0$, and spacetime becomes flat.
At $r\rightarrow\infty$, with the help of Eqs. (29), (39) we obtain the asymptotic value of the Ricci scalar
\begin{equation}
R=\kappa^2\beta\left(\frac{Q^4}{r^8}-\frac{7\beta Q^6}{2r^{12}}+ \frac{15\beta^2 Q^8}{r^{16}} +{\cal O}(r^{-20})\right).
\label{40}
\end{equation}

The metric function $f(r)$ for spherically symmetric line element (23), and the mass function $M(r)$ read
\begin{equation}
f(r)=1-\frac{2GM(r)}{r},~~~~M(r)=m+\int_0^r\rho_E(r)r^2dr=m+m_E-\int^\infty_r\rho_E(r)r^2dr,
\label{41}
\end{equation}
where $m_E=\int_0^\infty\rho_E(r)r^2dr$ is the electrostatic mass of the black hole and $m$ is the mass of the matter. To have the term containing
the mass $m$ in Eqs. (41), one has to introduce the Lagrangian density of the matter in Eq. (20). Using the expression for the energy density (7)
and Eq. (25) we find the mass function
\begin{equation}
M(r)=m+\frac{Q}{2}\int_0^r\frac{E(1+\beta E^2)dr}{1+\beta E^2/2}.
\label{42}
\end{equation}
where $E(r)$ is given by Eq. (25). From Eqs. (25) and (42) we calculate the electrostatic mass of the black hole
\begin{equation}
m_E=M(\infty)=\frac{1.5654 Q^{3/2}}{\beta^{1/4}}.
\label{43}
\end{equation}
The mass of the black hole has here the electromagnetic origin.
From Eqs. (29) and (42) we obtain the asymptotic value of the mass function at $r\rightarrow \infty$
\begin{equation}
M(r)=m+m_E-\frac{ Q^{2}}{2r}+\frac{\beta Q^4}{20r^5}-\frac{5\beta^2Q^{6}}{72r^9}+{\cal O}(r^{-13}).
\label{44}
\end{equation}
From Eqs. (41) and (44) one finds the metric function
\begin{equation}
f(r)=1-\frac{2G(m+m_E)}{r}+\frac{G Q^{2}}{r^2}-\beta\left(\frac{GQ^4}{10r^6}-\frac{5\beta GQ^6}{36r^{10}}+{\cal O}(r^{-14})\right).
\label{45}
\end{equation}
The first three terms in Eq. (45) are the RN solution and the last terms, containing the $\beta$, give corrections. At $Q=0$ in Eq. (45), we obtain the Schwarzschild solution.
At $r\rightarrow\infty$ the spacetime asymptotically approaches to the Minkowski spacetime (flat), and at $\beta=0$  one comes to the
Maxwell electrodynamics and solution (45) becomes the RN solution.
Some models of NLED \cite{Breton}, \cite{Hendi}, \cite{Kruglov6}, \cite{Kruglov8} also result in asymptotic RN black hole solution with
corrections. It follows from Eq. (45) that corrections to the RN solution change the event horizon. This also takes place for other NLED models
coupled to GR \cite{Breton}-\cite{Kruglov8}.

\section{Conclusion}

It should be mentioned that WEC, which means that the energy density as measured by local observer must be positive, imposes constrains on the
Lagrangian density. DEC includes WEC and requires that principal pressures are less than the energy density, which guarantees that sound speed
is less than the light speed. SEC defines the gravitational acceleration.
We have shown that in our model of NLED the energy conditions, WEC, DEC, and SEC, are satisfied.
The exact solution for the electric field of charged point-like objects possessing spherical symmetry have obtained and we found
the correction in the order of $r^{-6}$ to the Coulomb law. It should be noted that such corrections to Coulomb's law at $r\rightarrow\infty$ are different from corrections due to massive photons described by Yukawa potential.
The electric-magnetic duality transformations were studied. It was demonstrated that electric-magnetic duality between $P$ and ${\cal F}$ frames
is broken. We analyzed our NLED coupled to gravitational field and calculated the Ricci scalar asymptotic at $r\rightarrow \infty$ that is in the
order of $r^{-8}$. The electrostatic mass of the black hole expressed via the parameter $\beta$ and the electric charge was obtained.
We found the asymptotic of the metric function at $r\rightarrow\infty$ and corrections to the Reissner-Nordstr\"om solution.

\end{document}